\documentclass[12pt]{article}
\usepackage{geometry}
\usepackage{color}
\usepackage{xcolor}
\usepackage{a4}
\usepackage{graphicx}
\usepackage{epsf}
\usepackage{amsmath}
\usepackage{amssymb}
\usepackage{cite}
\newcommand{\be}{\begin{equation}}
\newcommand{\ee}{\end{equation}}

\newcommand{\Rmnum}[1]{\expandafter\@slowromancap\romannumeral #1@}
\newcommand{\bea}{\begin{eqnarray}}
\newcommand{\eea}{\end{eqnarray}}
\newcommand{\gae}{\lower 2pt \hbox{$\, \buildrel {\scriptstyle >}\over {\scriptstyle\sim}\,$}}

\begin{document}
\def\C{{\mathbb{C}}}
\def\R{{\mathbb{R}}}
\def\s{{\mathbb{S}}}
\def\T{{\mathbb{T}}}
\def\Z{{\mathbb{Z}}}
\def\W{{\mathbb{W}}}
\def\Bbb{\mathbb}
\def\BZ{\Bbb Z} \def\BR{\Bbb R}
\def\BW{\Bbb W}
\def\BM{\Bbb M}
\def\BC{\Bbb C} \def\BP{\Bbb P}
\def\CP{\BC\BP}
\begin{titlepage}
\title{Galactic space-times in modified theories of gravity} \author{} 
\date{
Dipanjan Dey, Kaushik Bhattacharya, Tapobrata Sarkar 
\thanks{\noindent 
E-mail:~ deydip, kaushikb, tapo @iitk.ac.in} 
\vskip0.4cm 
{\sl Department of Physics, \\ 
Indian Institute of Technology,\\ 
Kanpur 208016, \\ 
India}} 
\maketitle 
 
\abstract{We study Bertrand space-times (BSTs), which have been proposed as viable models of space-times seeded by galactic dark matter, in modified theories of gravity. 
We first critically examine the issue of galactic rotation curves in General Relativity, and establish the usefulness of BSTs to fit experimental data in this context. 
We then study BSTs in metric $f(R)$ gravity and in Brans-Dicke theories. For the former, the nature of the Newtonian potential is established, and we also compute the 
effective equation of state and show that it can provide good fits to some recent experimental results. For the latter, we calculate the Brans-Dicke scalar
analytically in some limits and numerically in general, and find interesting constraints on the parameters of the theory. Our results provide evidence for the 
physical nature of Bertrand space-times in modified theories of gravity.} 

\end{titlepage}
\section{Introduction}

Galactic dark matter has been one of the most intensely researched topics over the past several decades. Various models of dark matter that seek to establish the nature
of galactic dynamics \cite{BT} have been proposed and successfully tested with experimental results, although much still needs to be explored.
The role of Einstein's General Relativity (GR) in the study of galactic dynamics also has a long history (see, e.g \cite{sayan}, \cite{roberts}), although it is
fair to say that this is not a very popular approach among astrophysicists. This is possibly because of two reasons. Firstly, it is commonly believed
that at galactic length scales, the dynamics of celestial objects is necessarily Newtonian, and secondly there are various subtle issues regarding 
observers and measurements in GR which make practical applications of the theory to galactic dynamics somewhat complicated. 

The purpose of the present paper is to critically analyze some issues related to the application of GR and extended theories of gravity \footnote{The literature on the
subject is vast, and we refer the reader to the standard references \cite{Odintsov}, \cite{Sotiriou}, \cite{Capo}, \cite{BDbook} on the subject.} to galactic astrophysics. 
In particular, we focus on a class of space-time models proposed as viable models of galactic dark matter in \cite{dbs1}, \cite{dbs2}, which were originally discovered
by Perlick \cite{perlick} and called Bertrand space-times (BSTs). In these works, it was shown that in the framework of GR, BSTs can provide excellent fits to experimental data on galactic 
rotation curves. It was further established that these models can also accommodate observational results on gravitational lensing from galaxies and galaxy clusters. 

The main motivation for proposing BSTs as a viable phenomenological model of galactic dark matter (applicable maximally to low surface brightness galaxies) is that 
stars in the disc or halo regions of such galaxies move in closed stable orbits, at least to a good approximation. This geometric property of the orbits of stars is, by construction,
captured by BSTs in GR. One can thus think of BSTs as sourced by galactic dark matter. 
If BSTs are thought of as viable space-time models of dark matter, then we naturally move out of the paradigm of weak gravity. Although this new approach has not been 
very popular in existing literature, such a scenario cannot be ruled out, as experimental data can be well fitted within this framework \cite{dbs1},\cite{dbs2}. It is therefore
important and instructive to extend the analysis on such theories, in particular in the framework of modified theories of gravity, and this is 
one of the tasks that we undertake in this work. 

This paper is organized as follows. In the first section, we
examine various issues relating to galactic rotation curves in the GR framework. We point out that some standard definitions of the same used in the literature may not
be very useful in practise, as these cannot be realized in experimental situations. We establish that an alternative phenomenological definition of the circular velocity in GR in the context 
of BSTs may be more effective in galactic scenarios. Next, we move on to consider BSTs in extended theories of gravity. In section 3, we first ask if scalar fields can seed
a BST, and show that the answer is in the negative. Next, in section 4, we consider BSTs in the metric $f(R)$ gravity paradigm, and study various aspects of the same. In particular,
we show in a Newtonian context how the galactic potential is non-trivially modified in $f(R)$ BST theories. 
Finally, in section 5, BSTs are considered from the point of view of Brans-Dicke theories, and we obtain constraints on the Brans-Dicke parameter. Finally, section 6
ends this work with our conclusions and some prospective issues of future interest. 

\section{Galactic rotation curves and BSTs}

In General Relativity, one commonly uses the Schwarzschild metric to model galactic systems. This implies that, by assumption, gravity is weak, excepting
for regions close to the central singularity. To recapitulate some textbook numbers, we start with the Schwarschild metric
\begin{equation}
ds^2 = -c^2\left(1-\frac{2G_NM}{c^2r}\right)dt^2 + \frac{dr^2}{\left(1-\frac{2G_NM}{c^2r}\right)} + r^2d\Omega^2,
\label{Sch}
\end{equation}
with $d\Omega^2 = d\theta^2 + {\rm sin}^2\theta d\phi^2$ being the standard metric on the unit $2$-sphere, $c$ is the speed of light, and $G_N$ is the Newton's constant. 
Assume that the central point mass is $\sim 10^8 M_{\odot}$, which is
of the order of the mass of a typical galaxy like NGC4395. Then, using $G_N=4.3\times 10^{-3}~{\rm pc}M_{\odot}^{-1} ({\rm Km/sec})^2$, the Schwarzschild radius is at 
$r_s \sim 10^{-5}~{\rm pc}$. By a conservative estimate, if we assume gravity effects to be very small from $r \gae 10^2 r_s$, this would imply that GR effects may be negligible 
from $r \sim 10^{-3}~{\rm pc}$, while the radius of the galaxy is $\sim 1~{\rm Kpc}$. Hence, gravity is essentially Newtonian at galactic scales, and GR effects can safely be taken to be a small
correction to a Newtonian picture. While this model serves as the basis of an enormously successful theory of galactic dynamics, it is fair to say that it has its limitations when viewed
from the framework of GR. This is mainly because 
the Schwarzschild solution is a vacuum solution of Einstein gravity, and thus may not be very effective in describing dark matter dominated galaxies, given the fact that dark matter does 
affect celestial dynamics even away from the galactic centre. 

Another interesting possibility is to model galactic dynamics by other solutions of GR, which are not vacuum solutions. A standard approach \cite{sayan} is to write a galactic metric 
\begin{equation}
ds^2 = - e^{2\Phi(r)}c^2dt^2 + e^{2\lambda(r)}dr^2 + r^2d\Omega^2.
\label{kar}
\end{equation}
Here, $c$ is the speed of light, and the conserved energy $E$ and angular momentum $L$ (per unit mass) and their relation for circular orbits are given by 
\begin{equation}
E = c^2e^{2\Phi(r)}{\dot t},~~~L = r^2{\dot \phi},~~~E^2 = c^2e^{2\Phi}\left(c^2 + \frac{L^2}{r^2}\right),
\end{equation}
the dot denoting a derivative with respect to the proper time. For circular orbits, $\Gamma$ and $h$ are independent of the radial coordinate. However in practical situations, the orbits
may not be strictly circular, and hence $h$ might depend on the radial coordinate. In this situation, if $e^{2\Phi(r)} \sim 1$, and $L/r = {\rm v}_{\rm circ}$ is a constant where 
${\rm v}_{\rm circ}$ is  the circular velocity, then it can be shown that \cite{sayan}
\begin{equation}
e^{2\Phi(r)} = e^{-2{\rm v}_{\rm circ}^2/c^2}\left(\frac{r}{R}\right)^{2{\rm v}_{\rm circ}^2/c^2},
\end{equation}
that is, an explicit form of the $tt$ component of the metric is obtained from observational constraints, and in appropriate limits, this weak field analysis agrees with a corresponding
analysis with the Schwarzschild black hole. Note that the definition of the circular velocity here is ${\rm v_{circ}}=L/r = r{\dot \phi}$. We will record some observations here. Strictly
speaking, the meaning of circular velocity in a GR framework is somewhat ambiguous. A popular definition that has been used in the literature (see, e.g \cite{matos}) is 
${\rm v}_{\rm circ}=c\sqrt{rg_{tt}'/(2g_{tt})}$ (the prime denoting a derivative with respect to the radial coordinate $r$) where $g_{tt}$ is the $tt$ component of a metric with a generic form
\begin{equation}
ds^2 = g_{tt}(r)dt^2 + g_{rr}(r)dr^2 + r^2d\Omega^2.
\label{gen}
\end{equation}
There have been claims in the literature that this is a good definition of a circular velocity as measured by an observer at infinity. That this is not so is clear from the following 
elementary arguments \cite{hartle}. 
First note that for the general metric of Eq.(\ref{gen}), motion on the equatorial plane $\theta = \pi/2$ is described by the equation 
\begin{equation}
{\dot r}^2 + V(r) = 0,~~~V(r) = \frac{1}{g_{rr}(r)}\left[\frac{E^2}{g_{tt}(r)} + \frac{L^2}{r^2} + c^2\right],
\label{genmotion}
\end{equation}
where $E$ and $L$ are as before the conserved energy and angular momentum respectively, per unit mass. 
To measure circular speed (for recent work on the topic, see \cite{Phuc1}, \cite{Phuc2}), we need an inertial observer who uses a tetrad basis to project the four-momentum of a particle onto his frame, and equates
this to a Lorentzian form of the energy. Specifically, this means that the stationary observer measures the energy of a particle of rest mass $m$ as (Eq.(7.53) of \cite{hartle}) :
\begin{equation}
-p^{\mu}U_{\mu} =\frac{mc^2}{\sqrt{1-\frac{{\rm v}_{\rm circ}^2}{c^2}}}
\label{circ1}
\end{equation}
Here $p^{\mu}$ is the four-momentum of the particle, $U^{\mu}$ is the four-velocity of the observer, satisfying $U^{\mu}U_{\mu}=-c^2$. For a stationary observer, this latter
fact implies that the only non-vanishing component of the observer's four-velocity is $U^0=c^2/\sqrt{-g_{tt}(r)}$. We use this in conjunction with the fact that the time component of the particle's
four velocity is related to the conserved energy per unit mass, and for the metric of Eq.(\ref{gen}) is given by 
\begin{equation}
{\dot t} = -\frac{E}{g_{tt}},~~~E = c\sqrt{\frac{2g_{tt}^2}{rg_{tt}' - 2g_{tt}}}~,
\label{circ2}
\end{equation}
where the second relation in Eq.(\ref{circ2}) is obtained by solving for $V(r)=0,~V'(r)=0$ with $V(r)$ given from Eq.(\ref{genmotion}). \footnote{These are solved at the radius of the
circular motion. By a slight abuse of notation, we denote this by $r$ as well.} Now using Eqs.(\ref{circ1}) and (\ref{circ2}), we obtain 
\begin{equation}
\frac{{\rm v}_{\rm circ}^2}{c^2}= 1 + \frac{c^2g_{tt}}{E^2} \implies {\rm v}_{\rm circ} = c\sqrt{\frac{r g_{tt}'}{2 g_{tt}}},
\end{equation}
which is a definition conventionally used in the literature. It should be clear from our analysis that this definition of the circular velocity necessarily implies an observer who is stationary 
at a given point in the orbit of the test particle and this definition may not be very useful
in practise, as it requires a series of stationary observers at each of the radii of the celestial objects undergoing circular motion. A further drawback of this definition of the circular
velocity is that for calculation purposes, one has to often assume that this is a constant, thereby missing out the variations of the circular velocity as a function of $r$.

An alternative possibility is to use a phenomenological definition for the circular velocity, $v_{\rm circ} = rd\phi/dt$. This is motivated from the fact that for asymptotically flat observers
in GR, the quantity $d\phi/dt$ makes sense as an angular speed of an object in circular motion measured by an observer at infinity, whose proper time coincides with the coordinate time.
For a Schwarzschild background for example, it is a well known result that $d\phi/dt \sim 1/r^3$, i.e has the same form as in non-relativistic Keplarian motion. For the metric of 
Eq.(\ref{gen}), a simple calculation tells us that 
\begin{equation}
v_{\rm circ} \equiv r\frac{d\phi}{dt} = \sqrt{-\frac{rg_{tt}'}{2}}.
\label{vcircular}
\end{equation}
Clearly, for a Schwarzschild solution, Eq.(\ref{vcircular}) implies that $v_{\rm circ} \sim 1/\sqrt{r}$, i.e will always have a power law falloff. Or, if we want to study cases when the circular velocity
is a constant, then this implies that $g_{tt} \sim {\rm ln}~r$, i.e we need to go beyond a Schwarzschild approximation, to a paradigm where gravity is not modeled by a central 
point mass singularity. This was the issue we discussed in the beginning of this section. 

It is important to ask whether one can model galactic dynamics using metrics in which gravity is not negligibly weak beyond the central region. One such situation was envisaged in
\cite{dbs1},\cite{dbs2} where galactic space-times were modeled by a Bertrand  space-time (BST) metric of the form 
\begin{equation}
ds^2 = -c^2\frac{dt^2}{D +\frac{\alpha}{r}} + \frac{dr^2}{\beta^2} + r^2d\Omega^2,
\label{type2a}
\end{equation}
where $D$, $\alpha$ and $\beta$ are real and positive. This arises from the work of Perlick \cite{perlick} who showed that such metrics 
admit stable circular orbits at each point (for related work in Special Relativity, see \cite{kb}).\footnote{In the language of Perlick \cite{perlick}, the metric
of Eq.(\ref{type2a}) is a special case of what he has called Bertrand space-times of Type II, there being another version of the metric that supports closed, stable orbits at each point,
called BSTs of type I. Since we will always be dealing with the metric of Eq.(\ref{type2a}) in this paper, we will simply call this metric as the BST.}
If these orbits are closed, then $\beta$ has to be a rational number. This is a reasonable assumption for a 
galactic metric, given that at least in the outer regions of a galaxy, stars are known to move in stable closed orbits to a good approximation. 
The metric of Eq.(\ref{type2a}) can 
be treated as a phenomenological model for a dark matter dominated galaxy, for a number of reasons. Firstly, it can be checked that the alternative definition of circular velocity as given in the last paragraph yields (restoring
factors of $c$),
\begin{eqnarray}
v_{\rm circ}(r) = c\frac{\sqrt{\frac{\alpha r}{D}}}{\sqrt{2}D\left(r + \frac{\alpha}{D}\right)}
\label{vel}
\end{eqnarray}
It can be further shown that the radius at which the circular velocity maximizes, and the value of the maximum circular velocity are given by 
\begin{equation}
r_s = \frac{\alpha}{D},~~~~v_{\rm circ}^{\rm max} = \frac{c}{2\sqrt{2}}\frac{1}{\sqrt{D}}.
\label{rs}
\end{equation} 
Thus, in principle, the values of $D$ and $\alpha$ can be estimated by comparison with existing data for $v_{\rm circ}^{\rm max}$ and the radial distance 
at which the circular velocity maximizes. For a number of dark matter dominated galaxies, this was shown to give excellent fits to experimental data. 
Secondly, it can be checked that in a Newtonian approximation, the density profile predicted from Eq.(\ref{vel}) matches with the standard Navarro-Frenk-White (NFW) 
profile \cite{nfwprof} in the flat region of the rotation curves and the Hernquist \cite{hernquist} profile in general. 

The underlying reason for the metric of Eq.(\ref{type2a}) to match with data which are usually obtained from Newtonian physics can be stated as follows. 
If we substitute the metric of Eq.(\ref{type2a}) in Eq.(\ref{genmotion}),
then we get 
\begin{equation}
{\dot r}^2 + V(r) = 0,~~~V(r) =\beta^2c^2 - \frac{\beta^2}{c^2}E^2\left(D + \frac{\alpha}{r}\right) + \frac{\beta^2L^2}{r^2}.
\label{newtonian}
\end{equation}
Hence, apart from constant terms and the usual centrifugal barrier (the last term of Eq.(\ref{newtonian})), the potential  has a Newtonian form. We thus expect that in the framework of GR,
the metric of Eq.(\ref{type2a}) will be useful for contrasting and studying results otherwise obtained in the Newtonian framework, and as alluded to before, $\alpha$ and $D$
provide us with two paramters that can be used to fit galactic rotation curves. The caveat in our analysis is that the space-time described by the metric of Eq.(\ref{type2a}) is
not asymptotically flat. Apart from having a conical defect, we have also not rescaled the time coordinate, so that the $t$ that appears in Eq.(\ref{vel}) is the coordinate
time, and cannot be equated to the proper time of an asymptotically flat observer at infinity. We have to live with this fact, but emphasize here that our model is phenomenological,
and in a GR framework, our definition of the circular speed is closer in spirit to the ones measured in experiments. 

A few comments on our analysis are appropriate at this stage. First, 
we point out here that it is possible to rescale the time coordinate, so that it matches with the proper time at infinity. However it should be clear from the preceding discussion 
that on doing this,  fitting with observational data for galactic rotation curves becomes difficult (for example, the maximum value of the circular velocity 
in this case becomes $c/(2\sqrt{2})$, an unrealistically large number). The resolution of this problem is to match an internal BST with an external 
Schwarzschild solution, as discussed in section 2 of \cite{dbs1}. However, this often results in the presence of a thin shell of matter near the matching radius, and in such situations,
analytical handle on the problem might pose problems. We will thus retain the dependence on the parameter $D$ in the metric of Eq.(\ref{type2a}). 

Further, the constant $\beta$ appearing in the metric of Eq.(\ref{type2a}) (which has to be positive, as per Perlick's original construction \cite{perlick}) is restricted $0<\beta<1$. 
This is in order to keep the energy density for BSTs positive (see Eq.(\ref{rho}) of section 3). If $\beta$ is set to unity, which corresponds to Keplarian orbits in the original 
construction of \cite{perlick} then we can avoid a conical defect, but the energy density vanishes for $r > 0$. This might be somewhat unrealistic in a galactic scenario, 
hence we do not consider this case here. Also, as pointed out in \cite{dbs2}, the energy momentum tensor for the metric of Eq.(\ref{type2a}) can be represented by a anisotropic two-fluid
model.  Solving for the constraints of this model typically rules out values of $\beta$ close to unity (see section 3 of \cite{dbs2}). 
Also note that as we have mentioned, values of $\alpha$ and $D$ characterizes individual galaxies 
in our model, but $\beta$ plays no role in such a classification. This is because the circular velocity of Eq.(\ref{vel}) is independent of $\beta$, and so is the 
mass of the galaxy obtained from a Newtonian approximation (see Eq.(11) of \cite{dbs2}). 

A second issue of importance is the interpretation of the radial coordinate $r$. The $dr^2$ term in Eq.(\ref{type2a}) comes with a $\beta^2$, indicating a conical defect, i.e a non-trivial
holonomy at the origin. The understanding of a collapse process that results in the metric of Eq.(\ref{type2a}) with a conical defect is beyond the scope of the present work, and for our purposes,
we think of $r$ as the radius of a sphere at fixed values of $r$. Since our analysis of circular orbits is valid for such fixed values of the radial coordinate, we refer to $r$ as a galactic
radius, for the purpose of comparing with experimental data. 

\begin{figure}[t!]
\begin{minipage}[b]{0.5\linewidth}
\centering
\includegraphics[width=2.7in,height=2.3in]{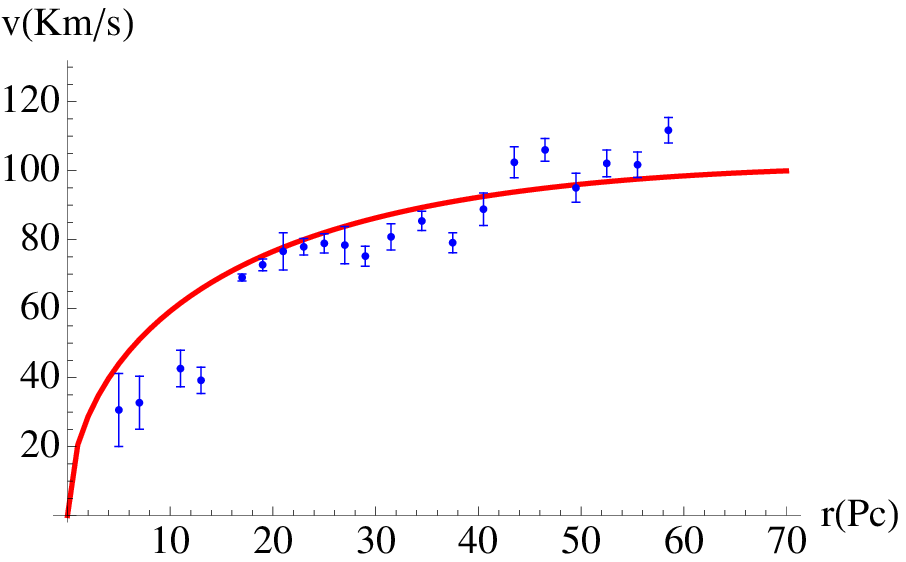}
\caption{Theoretical fit for circular velocity curve for the galaxy UGC 477 (solid red), compared with data (blue dots).}
\label{ugc477}
\end{minipage}
\hspace{0.2cm}
\begin{minipage}[b]{0.5\linewidth}
\centering
\includegraphics[width=2.7in,height=2.3in]{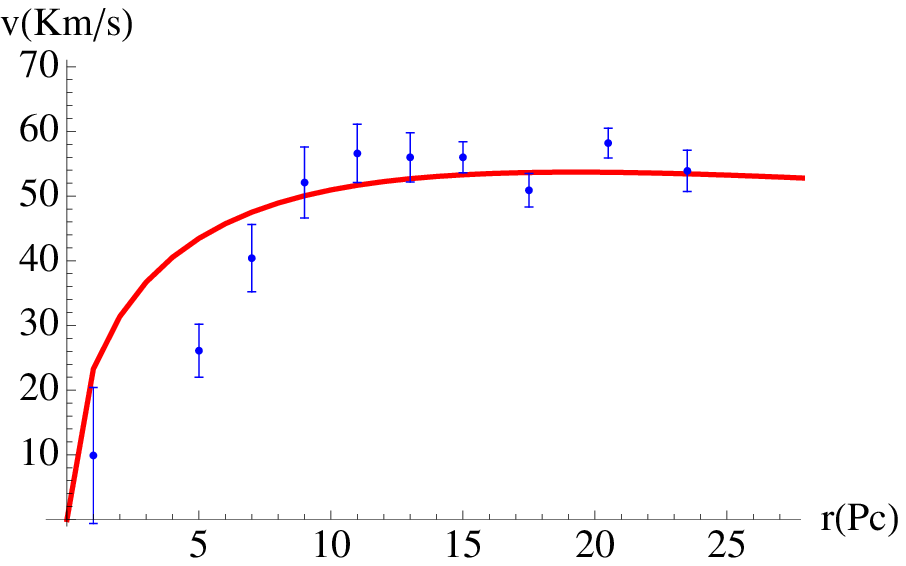}
\caption{Theoretical fit for circular velocity curve for the galaxy NGC 7137 (solid red), compared with data (blue dots).}
\label{ngc7137}
\end{minipage}
\end{figure}

Finally, one needs to understand the relationship between our proposed formula for the circular velocity of Eq.(\ref{vel}) with the corresponding spectroscopic result. Generically, the circular velocity
is measured by comparing the ratio of the frequencies of light emitted by a star and measured by an observer at infinity.  Within the framework of Schwarzschild gravity, the result
is given in standard textbooks, see e.g Eq.(11.25) of \cite{hartle}.  It is instructive to consider this in some detail. We begin with a generic static, spherically symmetric 
metric $ds^2 = g_{\mu\nu}dx^{\mu}dx^{\nu}$ with $\partial_t$ and $\partial_{\phi}$ being Killing vectors. Let $\omega_{\rm src}$ be the natural frequency of a photon emitted from a source
in circular motion on the equatorial plane ($\theta = \pi/2$), that is moving directly away from an observer at infinity. In that case, following the arguments of section 11.2 of
\cite{hartle}, it can be shown (with $g_{tt}$ negative) that if $\omega_{\rm obs}$ is the frequency of the photon measured by an observer at infinity, then 
\begin{equation}
\frac{\omega_{\rm obs}}{\omega_{\rm src}} = \frac{1}{\sqrt{-g_{tt}^{\rm obs}}}\frac{\sqrt{-g_{tt} - v_{\rm circ}^2}\sqrt{-g_{tt}}}{\sqrt{-g_{tt}} + v_{\rm circ}},
\label{freqratio}
\end{equation}
where on the right hand side, $v_{\rm circ} = rd\phi/dt$ is the circular velocity of the source, and $g_{tt}^{\rm obs}$ is the value of $g_{tt}$ measured at 
$r \to \infty$, i.e at the location of the observer. When $v_{\rm circ}=0$, this reduces to the familiar redshift factor. We now assume that $v_{\rm circ}^2$ is small
compared to $g_{tt}$. In BSTs, for a given range of $r$, the parameters $\alpha$ and $D$ can be
appropriately chosen so that this condition is satisfied. 
In this approximation, we expand the r.h.s of Eq.(\ref{freqratio}) upto second order in $v_{\rm circ}$ and obtain 
\begin{equation}
\frac{\omega_{\rm obs}}{\omega_{\rm src}} = \frac{\sqrt{-g_{tt}}}{\sqrt{-g_{tt}^{\rm obs}}} - \frac{v_{\rm circ}}{\sqrt{-g_{tt}^{\rm obs}}} + O(v_{\rm circ}^2).
\label{expand}
\end{equation}
The first term on the r.h.s of Eq.(\ref{expand}) gives the redshift factor and the second term estimates $v_{\rm circ}$ upto leading order. For the Schwarzschild
case, this reduces to the familiar Doppler shift formula for the frequency, as the red shift factor is close to unity if weak gravity is assumed. For BSTs, 
substituting Eq.(\ref{vel}) in Eq.(\ref{freqratio}), we find that
\begin{equation}
\frac{\omega_{\rm obs}}{\omega_{\rm src}} =
\frac{\sqrt{2} r \sqrt{\frac{\alpha }{D}+2 r}}{\sqrt{\frac{\alpha }{D}+r} \left(\sqrt{2} \sqrt{\frac{\alpha r}{D}}+2 \sqrt{r \left(\frac{\alpha }{D}+r\right)}\right)}
\label{spectro}
\end{equation}
The circular velocity calculated by the BST observer is then related to the frequency shift of Eq.(\ref{spectro}) minus the redshift factor (the first term
of Eq.(\ref{expand})). This is the spectroscopic interpretation of the circular velocity as measured by a BST observer at infinity. 
Specifically, choosing $\alpha/D = 1545.45$Pc with $D = 1.1\times 10^7$, we get a close fit for the roation curve of the galaxy
NGC4395. Similarly, with $\alpha/D = 1785.71$Pc with $D = 4.2\times 10^7$, a close fit to the rotation curve of the galaxy UGC1281 is obtained.
Comparative plots for these two cases appear in figure 2 and figure 3 of \cite{dbs2} respectively, to which we refer the reader. For the sake of completeness,
we have plotted, in fig.(\ref{ugc477}) and fig.(\ref{ngc7137}), the circular velocity curves for the galaxies UGC 477 and NGC 7137 from eq.(\ref{vel}) 
(solid red lines) and compared them with experimental data (blue dots) \cite{Data}. In the first case, we have made the choice $\alpha = 1.05 \times 10^8$ Pc and
$D = 1.1 \times 10^6$. In the second, we have chosen $\alpha = 7.5 \times 10^7$ with $D = 3.9 \times 10^6$. 

Let us summarize the main results of this section. Here, we have critically examined the delicate nature of the definition of circular velocity in a GR context, and showed that
conventional definitions that are often used in the literature may not be very useful from a practical point of view. We argued that a phenomenological definition given in 
\cite{dbs1},\cite{dbs2} might be more effective in comparing with experimental data, in the context of Bertrand space-times, which we propose as a viable space-time 
metric seeded by galactic dark matter. We also commented on the spectroscopic interpretation of the formula for the rotation curve for BSTs. 
Having thus established the usefulness of BSTs in the framework of GR in describing galactic dynamics, it is natural to investigate these in modified theories of gravity,
for example in $f(R)$ theories. This is the task that we undertake now. 

\section{BSTs in modified theories of gravity}

Before we start the main discussion about BSTs in a modified
theory of gravity, it is pertinent to ask if common matter, like a scalar
field or radiation can seed BSTs in GR. At first one may start with the
simple question: can real scalar fields seed a BST? If this is a possibility
then one can interpret the real scalar field to be a dark matter
field. Since BST metrics are static and spherically symmetric, we assume the solution $\phi \equiv \phi(r)$ and the
Lagrangian
\begin{eqnarray}
{\cal L} = \frac12 g_{\mu \nu}\partial^\mu \phi \partial^\nu \phi + V(\phi)\,,
\label{scl_lag}
\end{eqnarray}
where $V(\phi)$ is the scalar potential. Then, we require the following minimal conditions :
\begin{equation}
G_{\mu \nu}=\kappa T_{\mu \nu},~~~\Box~\phi(r)=V'(\phi(r)),
\end{equation}
where $\kappa = 8\pi G_N/c^4$, and the Einstein tensor $G_{\mu\nu}$ and the energy momentum 
tensor $T_{\mu\nu}$ are defined as 
\begin{equation}
G_{\mu\nu} = R_{\mu\nu} - \frac{1}{2}g_{\mu\nu}R,~~~T^{\mu \nu}=\partial^\mu \phi \partial^\nu \phi - g^{\mu \nu}{\cal L}.
\end{equation}

A well known example of such an Einstein Klein-Gordon system is the Janis-Newman-Winicour (JNW) \cite{jnw} space-time, which are singular space-times sourced by a scalar field, and given by the metric 
 \begin{equation}
ds^2_{\rm JNW} = -c^2\left(1-\frac{B}{r}\right)^\nu dt^2 + \frac{1}{\left(1-\frac{B}{r}\right)^\nu}dr^2 + r^2\left(1-\frac{B}{r}\right)^{1-\nu}d\Omega^2,
\label{a1}
\end{equation}
with $0<\nu<1$. The singularity of this space-time at $r=B$ is globally naked, and the solution of the scalar field is given by 
\begin{equation}
 \phi = \frac{q}{B\sqrt{4\pi}}\ln\left(1-\frac{B}{r}\right)
 \label{a2}
\end{equation}
where $q$ denotes its magnitude. The ADM mass $M$ is related to the parameters $B$ and $q$ by 
$B = 2\sqrt{q^2 + M^2}$. Also $\nu = 2M/B$, and in the limit $\nu \to 1$, i.e $q=0$, the Schwarzschild metric is recovered. 

On the other hand, the general form of the energy-momentum tensor and their relationship with the
energy density and principal pressures for BSTs are as follows \cite{kbs} :
\begin{eqnarray}
\rho(r) &=& -T^0_0 = \frac{1-\beta^2}{\kappa r^2}\,,
\label{rho}\\
p_r(r)  &=& T_1^1 = \frac{\beta^2(2\alpha + Dr)-(\alpha + Dr)}
{\kappa r^2(Dr + \alpha)}\,,
\label{pre}\\
p_\perp(r)  &=& T^2_2 = T^3_3 = \frac{\alpha\beta^2(\alpha - 2Dr)}
{4r^2\kappa(Dr+\alpha)^2}\,.
\label{pperpa}
\end{eqnarray}
Now, the general forms of $T^{00}$ and $T^{22}$ for the scalar field are
\begin{eqnarray}
T^{00} &=& - g^{00}\left[\frac{g_{11}}{2}\left(\frac{\partial\phi}{\partial r}\right)^2 + V(\phi)\right]\,,
\label{t00}\\
T^{22} &=& - g^{22}\left[\frac{g_{11}}{2}\left(\frac{\partial\phi}{\partial r}\right)^2 + V(\phi)\right]\,,
\label{t22}
\end{eqnarray}
from which we obtain
\begin{eqnarray}
\rho(r) &=& -T^0_0 = \left[\frac{g_{11}}{2}\left(\frac{\partial
    \phi}{\partial r}\right)^2 + V(\phi)\right]\,,
\label{sc_rho}\\
p_\perp (r) &=& T^2_2 = -\left[\frac{g_{11}}{2}\left(\frac{\partial
    \phi}{\partial r}\right)^2 + V(\phi)\right]\,,
\label{sc_press}
\end{eqnarray}
which implies $\rho = -p_\perp$, if $T^{\mu\nu}$ which seeds the space-time solely originates from a spherically symmetric real scalar field
distribution. It can be shown that this is indeed true for the JNW space-time. However, the energy-density and tangential pressure components of BST as given 
in Eqs.~(\ref{rho}) and (\ref{pperpa}) does not show $\rho = -p_\perp$ and consequently one can conclude that a BST cannot be seeded by a single real scalar field. An ideal
radiation field will also be unable to seed BSTs because its pressure
is isotropic. It seems that no candidate from known fluids is
useful enough for seeding  the BST in GR. Hence in GR one cannot avoid 
exotic fluids which might seed BSTs. 

The situation is more interesting in extended
theories of gravity, like $f(R)$ theories, where one may have
nontrivial space-time structure in absence of any matter. This can
happen because in these theories the curvature of space-time itself can
produce an effective energy density and pressure which can act as a
source of the space-time. Particularly, a nonstandard gravitational
theory like $f(R)$ theory is interesting in the case of BSTs because of
the properties of the Ricci scalar. For the BST of Eq.~(\ref{type2a}), the Ricci scalar turns out to be
\begin{equation}
R = \frac{\alpha^2(4-7\beta^2)+4Dr(1-\beta^2)(Dr+2\alpha)}{2r^2(Dr+\alpha)^2}\,,
\label{rk0}
\end{equation}
which diverges at $r \to 0$ where there is a naked singularity, and vanishes as $r \to \infty$. 

\begin{figure}[t!]
\begin{minipage}[b]{0.5\linewidth}
\centering
\includegraphics[width=2.7in,height=2.3in]{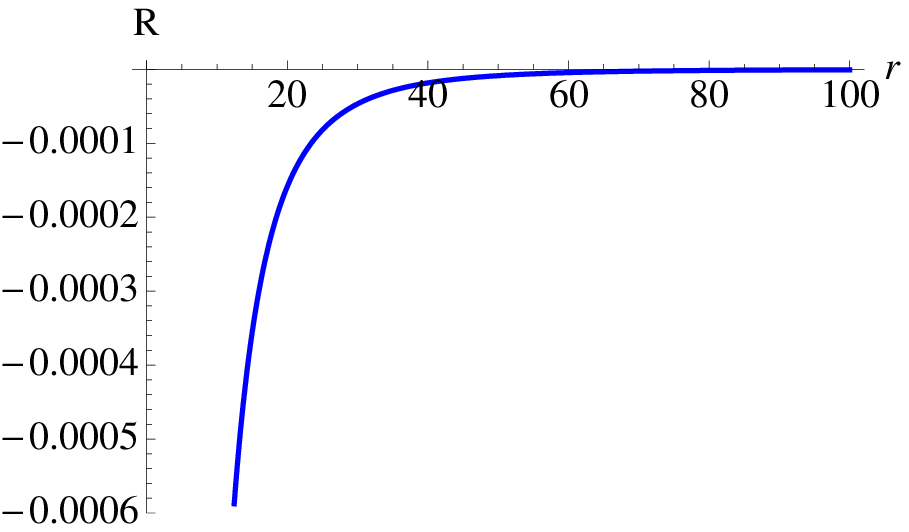}
\caption{Variation of the Ricci scalar $R$ with $r$ (in Kpc). Here $D=1.5\times 10^{5}$, $\beta=4/5$ and $\alpha=4.5\times 10^6$Kpc. 
$R$ is negative, and for this choice of parameters $R$ diverges when $r\to 0$ and becomes negligibly as $r$ increases to around $50$Kpc.}
\label{R0}
\end{minipage}
\hspace{0.2cm}
\begin{minipage}[b]{0.5\linewidth}
\centering
\includegraphics[width=2.7in,height=2.3in]{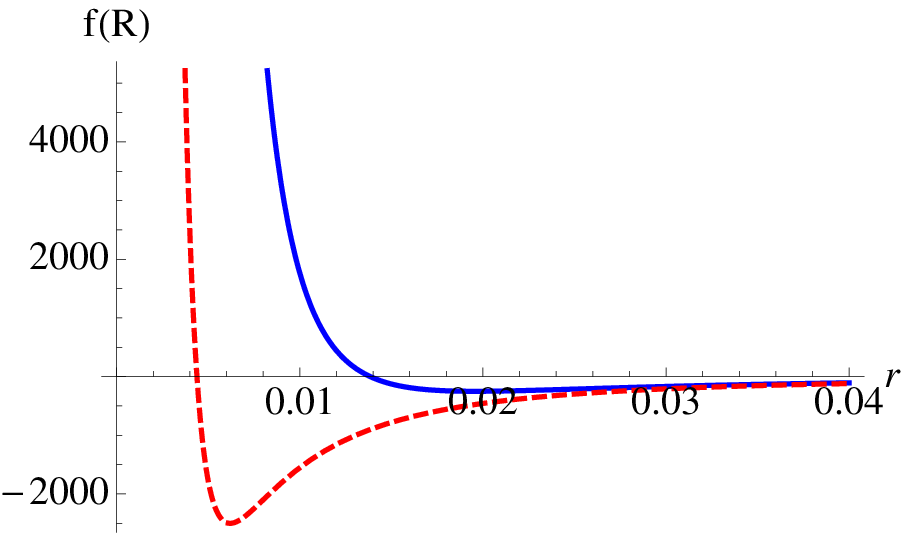}
\caption{Plot of $f(R)$ versus the radial coordinate $r$ for  $D=1.5\times 10^{5}$, $\beta=4/5$ and $\alpha=4.5\times 10^6$Kpc. 
$\lambda$ (in units of inverse length squared) equals $10^{-3}$ for the solid blue curve and $10^{-4}$ for the dashed red curve.}
\label{frs}
\end{minipage}
\end{figure}

If one looks at the variation of $R$ with respect to the radial
coordinate distance $r$, as shown in Fig.(\ref{R0}), it becomes clear
that the Ricci scalar diverges near the center and becomes negligible
$\sim 40$Kpc, for  $D=1.5\times 10^{5}$, $\beta=.8$ and $\alpha=4.5\times 10^6$Kpc. Because the Ricci scalar increases in magnitude
unboundedly very near the central singularity, it may happen that the
theory of gravity itself is modified near the center. The simplest
choice of an $f(R)$ where the corrections to GR becomes dominant when
the Ricci scalar starts to grow unboundedly is
\begin{equation}
f(R) = R + \lambda R^2,
\label{frexp}
\end{equation}
where $\lambda$ is a dimensionful parameter (of dimension inverse squared length, since the Ricci scalar has dimension length squared) and it sets the length
scale at which the correction term $\lambda R^2$ starts to contribute. Fig.(\ref{frs}) shows the effect of
$\lambda$ on the form of $f(R)$ where we have taken the same parameter choices as in Fig.(\ref{R0}). The solid blue and dashed red curves here
correspond to $\lambda = 10^{-3}$ and $10^{-4}$ respectively.\footnote{We will choose a positive sign for $\lambda$. This is dictated by the fact that a negative $\lambda$ seems to
render BSTs in $f(R)$ gravity unphysical. This will be explained in more details in the next section.} The absolute minima of $f(R)$ has
shifted more towards the centre (singularity) for the lower value of
$\lambda$. Although there can be generally many forms of $f(R)$ which
one may choose, the most general being a polynomial $f(R)$ with all
higher powers of $R$ appearing explicitly, our choice of $f(R)$ is the
simplest one among these. Our choice of quadratic
gravity does not eradicate the singularity at the centre but it can
make $f(R)$ finitely large near the centre by decreasing the value of
$\lambda$.  Unfortunately the simple form of $f(R)$, as given in
Eq.~(\ref{frexp}) cannot produce a consistent theory of gravitation
for BST in absence of any hydrodynamic matter and consequently we
require some form of matter to seed a BST even in an $f(R)$ theory. The
modified gravity solution of BST is presented in the next section of
this paper.

To summarise, in this section we have shown that a real scalar field or a radiation field cannot seed a Bertrand space-time. 
We have also motivated the fact that BSTs might be interesting to investigate in the context of modified theories of
gravity, like $f(R)$ theories. 
Before moving on to the next section, let us make some observations
about BSTs in modified theories of gravity.  From its
inception it was observed that $f(R)$ theories have an
interesting relationship with scalar-tensor theories of gravity. Any
arbitrary $f(R)$ theory can also be described by an
equivalent scalar-tensor gravitational theory, in particular the Brans-Dicke theory, 
in the Jordan frame when the Brans-Dicke parameter is set to
zero. To do so one does not require to do a conformal transformation.
Now, one may apply a conformal transformation on the metric and suitably
rescale the scalar field, in the Jordan frame where the $f(R)$ theory
is defined, and recast the whole theory in the Einstein frame, as commonly done
in cosmology. 

The Einstein frame description of the original $f(R)$ theory is equivalent
to a theory which involves Einstein gravity, in the presence of a
minimally coupled real scalar field, and a hydrodynamic fluid, in case
of non-vacuum $f(R)$ solutions. In cosmology, one often uses the
Einstein frame description of the dynamics of $f(R)$ theories in
the Jordan frame, perhaps the most famous example of this method was
applied by Starobinsky \cite{Starobinsky:1980te} in his theory of inflation where
he chose an $f(R)$ whose form is that given
in Eq.~(\ref{frexp}). In cosmology, the method of analyzing the FRW
solution of a $f(R)$ theory in the Einstein frame succeeds because
under a conformal transformation a FRW solution remains a FRW
solution. In the case of BSTs, this formalism of tracking the
gravitational behaviour of $f(R)$ theories in the Einstein frame does
not work due to the simple fact that a BST does not transform to another
BST with some redefined parameters under a conformal
transformation. As a consequence of this, BSTs in a
general Brans-Dicke theory (with a non-zero Brans-Dicke parameter),
$f(R)$ theory and conventional GR cannot be connected in any
mathematical form and have to be separately analyzed. In a
previous publication \cite{dbs2} the analysis of BST solutions in GR
was presented, in this paper we present the solutions in $f(R)$ theory
and in Brans-Dicke theory.

In the next section we will also show that in $f(R)$ description of
BSTs, one needs matter and from our conjecture that BSTs can serve as galactic space-times, this
matter can be interpreted as the ubiquitous dark matter.

\section{BSTs in the metric $f(R)$ gravity paradigm}
\label{bst2fr}

In  metric $f(R)$ gravity the action functional is
\begin{eqnarray}
S = \frac{1}{2\kappa} \int d^4 x \,\sqrt{-g}\,f(R) + S_{\rm Mat}\,,
\label{fract}
\end{eqnarray}
where $f(R)$ is a function of the Ricci scalar and $S_{\rm Mat}$ is the action for the matter fields. By varying the metric one arrives at
the equation 
\begin{eqnarray}
F(R)R_{\mu \nu}-\frac12f(R)g_{\mu\nu}-[\nabla_\mu \nabla_\nu - g_{\mu\nu}\Box]F(R)
=\kappa T_{\mu\nu}^{\rm Mat}\,,
\label{freqn}
\end{eqnarray}
where we have denoted
\begin{eqnarray}
F(R) \equiv f^{\prime}(R)\,,
\label{capf}
\end{eqnarray}
and here and in sequel, the primes will denote the differentiation with respect to $R$, and $\nabla_\mu$ designates  covariant derivatives with $\Box=\nabla_\mu
\nabla^\mu$. Here $T_{\mu \nu}^{\rm Mat}$ is the conventional energy-momentum tensor due to the matter fields. Using the form of the Einstein tensor, 
$G_{\mu\nu}\equiv R_{\mu\nu}- \frac12 g_{\mu \nu}R$, one can write Eq.~(\ref{freqn}) in a way which is similar to the Einstein equation in GR. 

Importantly, we assume here that BSTs are valid solutions of Eq.(\ref{freqn}), and check the viability of this assumption. In our model, the Einstein tensor will be 
calculated from the metric of Eq.(\ref{type2a}). Our analysis here should be contrasted with the more standard approaches in the literature \cite{Capo2} where 
modified gravity theories are used in the weak field limit to construct a modified gravitational potential which is then constrained by fitting with galactic rotation
curves. In particular, in \cite{Capo2}, \cite{Capo3}, the authors construct metric solutions of $f(R)$ theory by assuming some general forms of $f(R)$ and weak gravity. In
these references, the authors point out the interesting fact that flat velocity rotation curves for galaxies can be obtained without any explicit need of dark
matter, in an $f(R)$ gravity paradigm. It is to be noted that in our work we do not solve for the metric using $f(R)$ or Brans-Dicke theory
and apriori used the Bertrand spacetime as a solution for these.
In particular, assuming that BSTs are solutions of $f(R)$ gravity, that the galactic rotation curves are the same as the ones
discussed in the previous section. 

With the BST solution, we then compute the matter density and the principal pressures in $f(R)$ gravity. 
The Einstein like equation in metric $f(R)$ gravity is (with $\kappa = 1$) :
\begin{eqnarray}
G_{\mu \nu}= {\cal T}_{\mu \nu}\,,
\label{einslike}
\end{eqnarray}
where the effective energy momentum tensor is 
\begin{eqnarray}
{\cal T}_{\mu \nu} \equiv \frac{1}{F(R)}\left[T^{\rm Mat}_{\mu \nu} + 
\frac12(f(R) - RF(R))g_{\mu \nu} + (\nabla_\mu \nabla_\nu -
g_{\mu\nu}\Box)F(R) \right]\,,
\label{curlyt}
\end{eqnarray}
The first term within brackets on the right hand side of Eq.(\ref{curlyt}) is the matter contribution to the effective 
energy momentum tensor, and the rest is interpreted as the contribution due to curvature. 
Using the expression for $R$ of Eq.(\ref{rk0}), and a given form for $f(R)$, we can calculate the matter part of the energy momentum tensor $T^{\rm mat}_{\mu\nu}$ and 
this is what we focus on for the moment. This is calculated by using Eq.(\ref{curlyt}) : 
the effective energy density in the present case is $\rho^{\rm eff}=-{\cal T}^0_0$ and this equals the expression in Eq.(\ref{rho}). 
Similarly, the principal pressures can be calculated
using the diagonal terms in ${\cal T}^i_j$, and coincide with Eqs.(\ref{pre}) and (\ref{pperpa}). 

For further analysis, we find it convenient to choose the specific form of $f(R)$, as given in Eq.~(\ref{frexp}). We now present a few comments regarding the sign of $\lambda$. 
In this paper, we will take $\lambda$ to be a dimensionful small parameter, which is positive definite. We note here that in general, the sign of $\lambda$ may be constrained from 
a weak field analysis \cite{Capo},\cite{capo1}. However, here we do not pursue this line of approach due to the following reason. For a Schwarzschild type metric, the weak field analysis approximates 
$g_{tt} \simeq -(1 + \frac{2\Phi(r)}{c^2})$, with $\Phi(r)$ being the Newtonian potential. From the metric of Eq.(\ref{type2a}), such an approximation would amount to setting $r \gg r_s$, where  
$r_s = \alpha/D$ (see Eq.(\ref{rs})). However, in dark matter dominated
galaxies, it has been shown that to a good approximation, we can take the dark matter region of the galaxy to end at $r = r_s$ \cite{dbs1},\cite{dbs2}. The traditional weak field limit would hence be 
effective only very far from the galactic centre with no dark matter, and in a BST, this is not an interesting region to look at. Hence, the weak field analysis is less useful in our case.
The form of $f(R)$ that we have taken in Eq.~(\ref{frexp}) is phenomenological in nature, and  
the physical constraint of the positivity of the energy density dictates that we choose a positive sign of $\lambda$.\footnote{We will momentarily see that for the JNW space-time, a similar
physicality condition dictates that $\lambda$ is negative.} Choosing the negative sign gives rise to negative energy
densities, as can be checked, indicating an unphysical theory. 

We also set, in Eq.(\ref{type2a}), $\beta = 4/5$. Then from Eq.~(\ref{curlyt}), the
matter contribution to the energy-density for BSTs turn out to be
\begin{eqnarray}
\rho^{\rm Mat} = \frac{9}{25r^2} + \frac{6 \lambda  \left(43 \alpha ^4-165 D^4 r^4-660 \alpha  D^3 r^3+546 \alpha ^2 D^2 r^2+364 \alpha ^3 D r\right)}{625 r^4 (\alpha +D r)^4},
\label{rhoeff0}
\end{eqnarray}
where the second term can be interpreted as the energy density arising due to curvature effects, and vanish as $\lambda \to 0$. 
At this point, it is instructive to consider in some details the physics of Eq.(\ref{rhoeff0}). First, let us consider the GR case, i.e set $\lambda = 0$ in this equation. Then, we obtain
(as in Eq.(\ref{rho}), with $\kappa = 1$), $\rho^{\rm Mat} = 9/(25r^2)$, which is the density distribution of the  singular isothermal sphere. For this distribution, from a purely Newtonian perspective, 
the Poisson's equation $\nabla^2 \Phi = 4\pi G_N\rho$ is satisfied by $\Phi = (36/25)\pi G_N{\rm ln}(r)$ and from the relation (see e.g. Eq.(2.29) of \cite{BT})
\begin{equation}
r\frac{d\Phi}{dr} = v_{\rm circ}^2,
\label{bin}
\end{equation}
we get the well known result that the circular velocity is a constant, i.e $v_{\rm circ} = (36/25)\pi G_N$. On the other hand, our phenomenological definition of $v_{\rm circ}$ of Eq.(\ref{vel}) yields,
via Eq.(\ref{bin}),
\begin{equation}
\Phi = -\frac{\alpha c^2}{2 D (\alpha+D r)},
\end{equation}
and from Poisson's equation this gives rise to the Hernquist profile 
\begin{equation}
\rho = \frac{\alpha^2c^2}{4\pi G_N}\frac{1}{r(\alpha + Dr)^3}.
\end{equation}
In the framework of $f(R)$ gravity, we consider the matter density of Eq.(\ref{rhoeff0}). Of course, for metric $f(R)$ gravity, the Poisson equation is modified from its usual form, as is
known from a weak field analysis (see, e.g \cite{capo1}).  A rigorous analysis for the Poisson's equation in BSTs, in lines of \cite{capo1} will be presented elsewhere. Here we simply note that 
from a Newtonian perspective, from Eq.(\ref{rhoeff0}) we can derive a potential
\begin{equation}
\Phi = \frac{36}{25}\pi G_N {\rm ln}(r) + \frac{12\pi G_N\lambda}{625\alpha^2r^2(\alpha + Dr)^2}{\mathcal A},
\label{potfreq}
\end{equation}
where we have defined
\begin{eqnarray}
{\mathcal A} &=& \alpha \left(43 \alpha^3-1226 \alpha^2 D r-1989 \alpha D^2 r^2-960 D^3 r^3\right)\nonumber\\
&+&192 D r (2 \alpha+5 G r) (\alpha+D r)^2 \left[\ln (\alpha+D r)-\ln (r)\right].
\end{eqnarray}
\begin{figure}[t!]
\centering
\includegraphics[width=2.7in,height=2.3in]{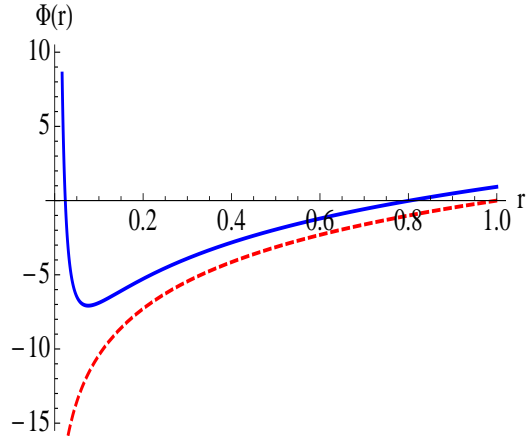}
\caption{Galactic potential as a function of the radial distance for BSTs in GR (dashed red) and $f(R)$ gravity (solid blue) (see text for details).}
\label{potfr}
\end{figure}
It is interesting to note the change in the nature of the potential in $f(R)$ gravity, compared to the GR case. In Fig.(\ref{potfr}) we contrast the two situations, where we have set
$\alpha = 10^5 {\rm Mpc}, D = 10^5, G_N = 1$.\footnote{In this section, the same values of $\alpha$ and $D$ will be chosen in sequel and we will not  mention this further.}
Here, the solid blue line is the potential of Eq.(\ref{potfreq}) with $\lambda = 10^{-3}$, while the dashed red line is the corresponding situation in GR, 
with $\lambda = 0$. We see that the effect of the curvature correction to the potential is to
modify it at small distances, where a minimum of the potential develops. This means that at this minimum, the circular velocity is zero, from Eq.(\ref{bin}), and does not exist below 
this distance. We emphasize that our results are only indicative and that we have resorted to a naive analysis in a Newtonian paradigm.
It should be interesting to explore this further. 

Now for the sake of completeness, we record the expression for the principal radial pressure due to matter, which we find to be
\begin{eqnarray}
p^{\rm Mat}_r &=& p_{1r} + \lambda p_{2r}~;\nonumber\\
p_{1r} &=& \frac{7 \alpha - 9Dr}{25 r^2 (\alpha +D r)},\nonumber\\
p_{2r} &=& -\frac{6  \left(-149 \alpha ^4+411 D^4 r^4+1644 \alpha  D^3 r^3+1442 \alpha ^2 D^2 r^2-20 \alpha^3 D r\right)}{625 r^4 (\alpha +D r)^4}
\label{pr}
\end{eqnarray}
and similarly, the matter contribution to the tangential pressures ($p^{\rm Mat}_2 = p^{\rm Mat}_3 \equiv p^{\rm Mat}_{\perp}$) are obtained as 
\begin{eqnarray}
p^{\rm Mat}_{\perp} &=& p_{1\perp} + \lambda p_{2\perp}~;\nonumber\\
p_{\perp 1} &=& \frac{4 \alpha  (\alpha - 2 D r )}{25 r^2 (\alpha +D r)^2}, \nonumber\\
p_{\perp 2} &=& -\frac{6 \left(101 \alpha ^4-411 D^4 r^4-1500 \alpha D^3 r^3+14 \alpha ^2 D^2 r^2+260 \alpha^3 D r\right)}{625 r^4 (\alpha +D r)^4}.
\label{pperp}
\end{eqnarray}

A few words about the energy conditions in BSTs in the framework of $f(R)$ theories is in order. First, we recapitulate some basic facts regarding these in GR (we will closely follow the discussion of \cite{poisson}).
In a locally flat tetrad basis, we assume that the energy momentum tensor can be decomposed as 
\begin{equation}
T^{\mu\nu} = \rho e^{\mu}_0e^{\nu}_0 + p_1e^{\mu}_1e^{\nu}_1 + p_2e^{\mu}_2e^{\nu}_2 + p_3e^{\mu}_3e^{\nu}_3
\label{tetrad}
\end{equation}
where we have the standard relation between the tetrads $e^{\mu}_{a}$, i.e $g_{\mu\nu}e^{\mu}_{a}e^{\nu}_{b} = \eta_{ab}$, with $\mu,\nu,\cdots$ denoting curved space indices and 
$a,b,\cdots$ are the flat space indices with metric $\eta_{ab}={\rm diag}(-1,1,1,1)$. Then, we have $\rho = -T^0_0$, $p_i = T^i_i~({\rm no ~ sum})$, $i=1,2,3$. The weak energy condition (WEC) is
then $T_{\mu\nu}u^{\mu}u^{\nu} \geq 0$ where $u^{\mu}$ is a future directed timelike vector. This boils down to the conditions $\rho \geq 0$, $\rho + p_i \geq 0$. The strong energy condition (SEC) is on the other
hand, a statement about the Ricci tensor, since it is given by the condition $(T_{\mu\nu} - \frac{1}{2}g_{\mu\nu}T)u^{\mu}u^{\nu} \geq 0$, with the Einstein's equations dictating that $T_{\mu\nu} - \frac{1}{2}g_{\mu\nu}T=R_{\mu\nu}$.
For $f(R)$ gravity, if we assume the SEC to be similarly defined, i.e $R_{\mu\nu}u^{\mu}u^{\nu} \geq 0$, and that the effective energy-momentum tensor ${\cal T}^{\mu\nu}$ of Eq.(\ref{curlyt})
admits the same decomposition as in Eq.(\ref{tetrad}), then we have $\rho^{\rm eff} + \sum_i p_i^{\rm eff} \geq 0$. Similarly, the WEC is given in $f(R)$ gravity by $\rho^{\rm eff} + p_i^{\rm eff} \geq 0$ (see, e.g \cite{santos}). 
That these are satisfied in our case follows from the fact that as mentioned in the beginning of this section, we have assumed that BSTs are solutions to Eq.(\ref{freqn}), and energy conditions for
BSTs of the form presented in Eq.(\ref{type2a}) have been established in \cite{kbs}. 
\begin{figure}[t!]
\begin{minipage}[b]{0.5\linewidth}
\centering
\includegraphics[width=2.7in,height=2.3in]{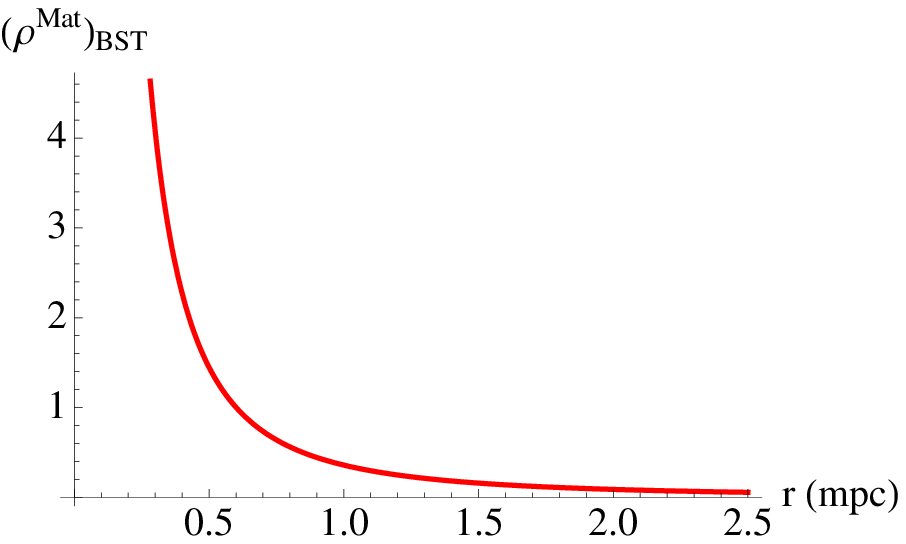}
\caption{Energy density of matter for BSTs in $f(R)$ gravity (see text for details).}
\label{bstrho}
\end{minipage}
\hspace{0.2cm}
\begin{minipage}[b]{0.5\linewidth}
\centering
\includegraphics[width=2.7in,height=2.3in]{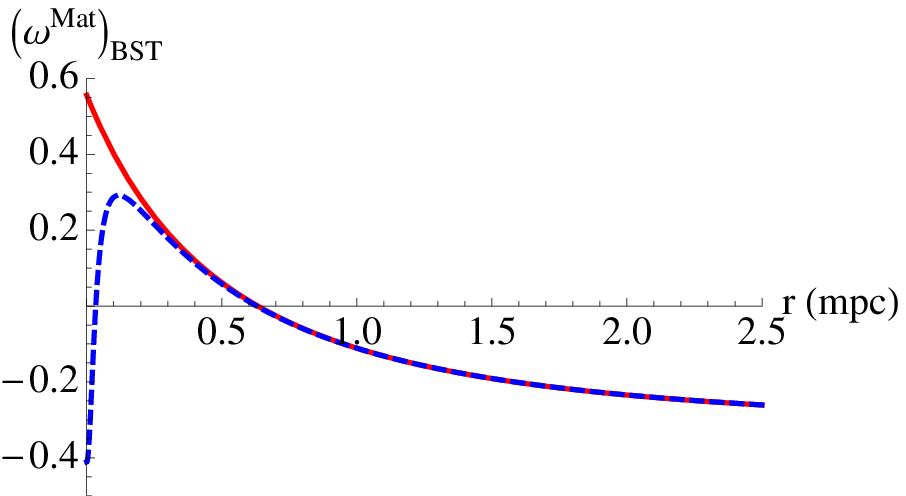}
\caption{Equation of state parameter for BSTs in $f(R)$ gravity (see text for details).}
\label{bstw1}
\end{minipage}
\end{figure}

It is reasonable to demand that the matter contribution to the energy density of Eq.(\ref{rhoeff0}) is positive definite. That this is so for BSTs is shown in Fig.(\ref{bstrho}), where we have plotted
$\rho^{\rm Mat}$ of Eq.(\ref{rhoeff0}) as a function of the radial coordinate, using $\lambda = 10^{-3}$. One can see that the matter contribution to the energy density remains positive for all values of $r$.\footnote{The 
curve for $\lambda = 0$ almost coincides with the the one shown for $r \geq 0.5$. Expectedly, they differ significantly for very small values of $r$, but this is not shown here.} Now, we make some comments
about the possible equation of state (EOS) of dark matter in our model. This topic has received some interest of late, following the work of \cite{sayan}, \cite{visser}. In \cite{serra}, \cite{sartoris}, the authors
computed an effective EOS parameter
\begin{equation}
\omega^{\rm Mat} = \frac{\sum_i p_i(r)}{3\rho(r)},
\end{equation}
from data on the weak lensing behavior and rotation curves. In particular, these authors measure $\omega^{\rm Mat}$ for the Coma Cluster and the CL0024 cluster which are galaxy clusters in which the dark matter
content is known to be $90 \%$ of the total matter. 

Such situations are ideal for BST models, where we can compute this quantity using Eqs.(\ref{rhoeff0}) - (\ref{pperp}). This is presented in Fig.(\ref{bstw1}). Here the solid red curve is for 
$\lambda = 0$ (i.e the GR case) and the dashed blue curve is for $f(R)$ gravity, where we have set $\lambda = 10^{-3}$. Expectedly, these are different for small radii and match for large values of the radius. 
While both the curves asymptote to $-\frac{1}{3}$ for very large $r$, the solid red curve ($\lambda = 0$) asymptotes to $5/9$ as $r\to 0$, the dashed blue curve asymptotes to $-0.41$ in this limit. 
The large $r$ behavior of the curves is of course reminiscent of the SEC satisfied by the matter contribution to the energy momentum tensor. We note here that the result presented in Fig.(\ref{bstw1}) is very similar
to the ones obtained in \cite{serra}, although the latter resuts were in the weak field limit. Our results are however at variance with those of \cite{sartoris}, where the authors obtain evidence for 
pressureless dark matter, i.e $\omega^{\rm Mat} \sim 0$. 
\begin{figure}[t!]
\begin{minipage}[b]{0.5\linewidth}
\centering
\includegraphics[width=2.7in,height=2.3in]{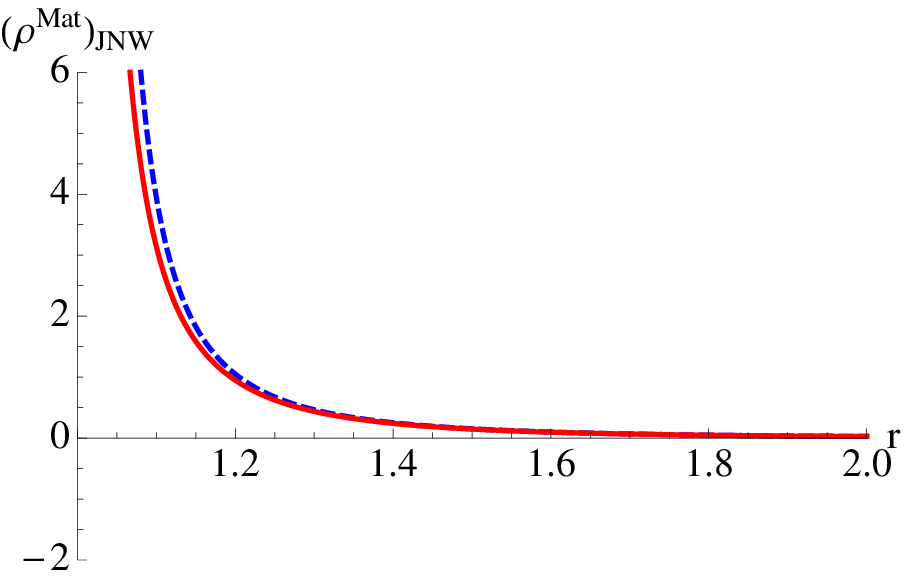}
\caption{Energy density of matter for JNW space-times in $f(R)$ gravity (see text for details).}
\label{jnwrho}
\end{minipage}
\hspace{0.2cm}
\begin{minipage}[b]{0.5\linewidth}
\centering
\includegraphics[width=2.7in,height=2.3in]{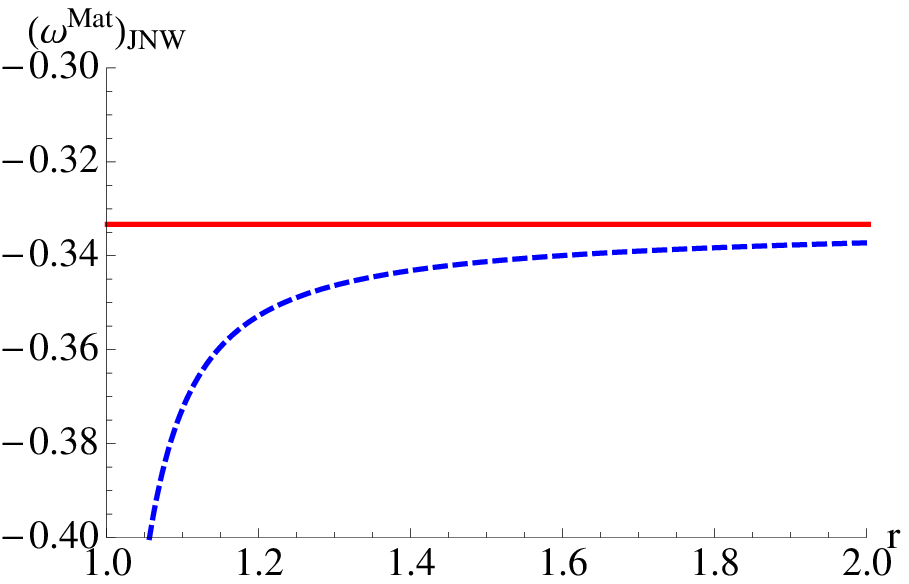}
\caption{EOS parameter for JNW space-times in $f(R)$ gravity (see text for details).}
\label{jnww}
\end{minipage}
\end{figure}

It is instructive to compare the BST result with that of the JNW naked singularity, with the metric given by Eq.(\ref{a1}). 
We assume that this metric is a valid solution to Eq.(\ref{freqn}). We will assume $\nu = 0.6$ and $B = 1$ without loss of generality. 
In Fig.(\ref{jnwrho}) and Fig.(\ref{jnww}), we show graphically the energy density due to matter and the effective equation of state with the dashed blue 
lines, where we have taken $f(R) = R + \lambda R^2$, with $\lambda = -10^{-3}$. 
The solid red lines are for the GR case, i.e $\lambda = 0$. In this case, we find that for positive values of $\lambda$, the 
matter contribution to the effective stress energy tensor becomes negative, and hence this is ruled out. 

\begin{figure}[t!]
\begin{minipage}[b]{0.5\linewidth}
\centering
\includegraphics[width=2.7in,height=2.3in]{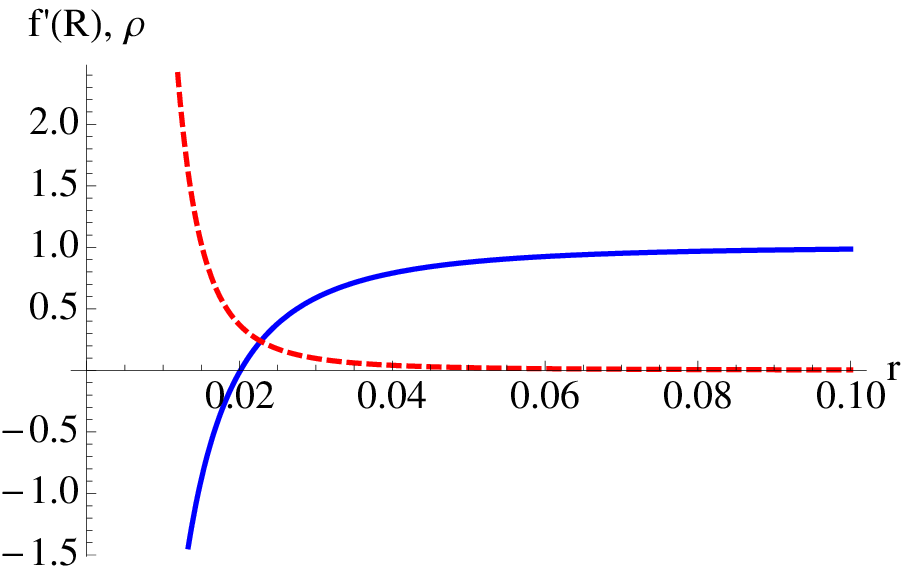}
\caption{$f'(R)$ (solid blue) and $\rho$ (dashed red) as a function of $r$ for $\lambda = 10^{-3}$ for BSTs (see text for details).}
\label{valid1}
\end{minipage}
\hspace{0.2cm}
\begin{minipage}[b]{0.5\linewidth}
\centering
\includegraphics[width=2.7in,height=2.3in]{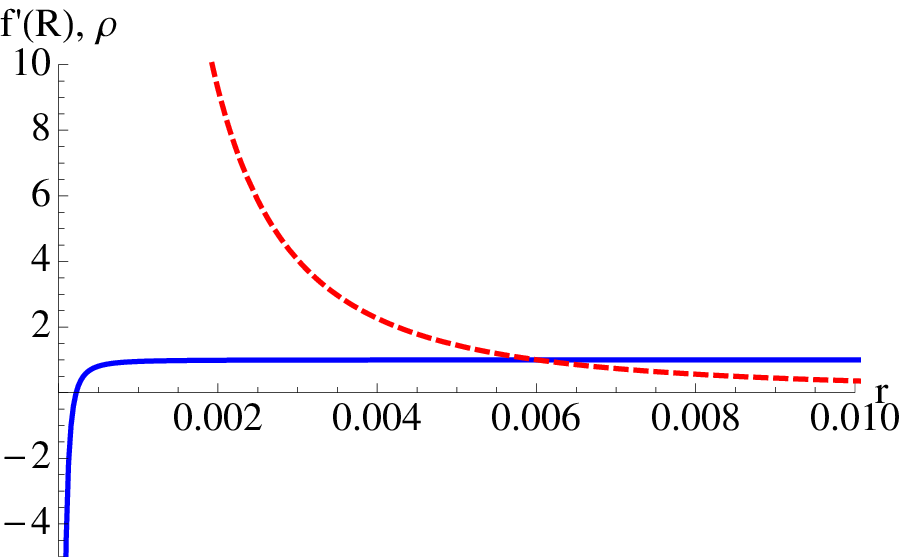}
\caption{$f'(R)$ (solid blue) and $\rho$ (dashed red) as a function of $r$ for $\lambda = 10^{-7}$ for BSTs (see text for details).}
\label{valid2}
\end{minipage}
\end{figure}

Before we end this section, we should point out a caveat in our analysis and its possible resolution. As is well known 
(see, e.g the review \cite{Sotiriou}), in realistic $f(R)$ theories, we require that $f'(R) > 0$, so that the effective gravitational constant $G_{\rm eft} = G/f'(R)$ is positive. 
This condition is required so that there are no ghost modes in a quantized version of the theory. If we assume 
the form of $f(R)$ in Eq.(\ref{frexp}), then it is clear that this condition will not be satisfied for small values of the radial coordinate, since the curvature diverges to negative 
infinity (see Eq.(\ref{rk0}) and Fig.(\ref{R0})). While this seems to be a problem, a possible resolution is to assume that the theory ceases to be valid for values of $r$ 
close to the central singularity. The lower cutoff for $r$ will depend on the chosen value of $\lambda$. In Figs.(\ref{valid1}) and (\ref{valid2}), we show the variation of  
$f'(R)$ for $\lambda = 10^{-3}$ and $10^{-7}$ respectively by the solid blue lines. The dashed red lines are the corresponding values of the density $\rho$. It is seen 
that as we decrease the value of $\lambda$, the region of invalidity of the condition $f'(R) >0$ becomes smaller. It is thus possible to choose a small value of $\lambda$ 
so that the lower cutoff of the theory is sufficiently close to the central singularity where anyway quantum effects might become important. There is a window of allowed 
values of $r$ for which $f'(R) < 1$ and this leads to a spatially varying Newton's constant, a situation that we will encounter in the context
of the BSTs in the Brans-Dicke theory which we now proceed to study. 

It is interesting to note that there can be some choices of $f(R)$
where one can demand that $f'(R) > 0$. As an example, if one chooses
$f(R)=R+\lambda_1 R^2 + \lambda_2 R^3$ where $\lambda_1$ and
$\lambda_2$ are constant parameters, then $f'(R)>0$ if $\lambda_1^2 <
3\lambda_2$. A different but reasonably simple form of $f(R)$ can be
$f(R)=(e^{\lambda_3 R}-1)/\lambda_3$ for which $f'(R)>0$ irrespective
of the value of the constant parameter $\lambda_3$. If one applies this forms of $f(R)$ to study the gravitational aspects of BST then it turns out 
that although $f'(R)$ can be made positive for all values of $r$, the energy density $\rho^{\rm Mat}$, becomes negative very close to $r=0$. 
In this cases one does not have any difficulty with the gravitational theory as such but the negative value of the matter energy density near the core of 
the galaxy shows that such a space-time cannot be seeded by any form of conventional matter. In such cases also one can proceed by demanding 
that the theory makes sense as long as $\rho^{\rm Mat}$ is positive. This discussion shows that it is very difficult to assure both $\rho^{\rm Mat} > 0$ and 
$f'(R) > 0$ for all radial distances as one of the two turns out to be negative very near the core. In this article we have preferred the positivity of the energy 
density over the positivity of the first derivative of $f(R)$ and consequently we do not discuss 
more on the gravitational theories resulting from the new forms of $f(R)$ discussed here. 

To summarize, the main results of this section are as follows. We have investigated here Bertrand space-times in the framework of metric $f(R)$ gravity, by taking it to 
be a solution of Eq.(\ref{freqn}). In this formalism, we calculated the
matter density and principle pressures. From the former, we obtained the Newtonian potentials that satisfy the Poisson's equation and contrasted the results obtained here with those in GR. 
We found that the potential shows an interesting deviation in $f(R)$ gravity. We further analyzed the energy conditions, and checked their validity. We then studied the effective equation of state
parameter in $f(R)$ gravity and showed that this is close to some of the existing results \cite{serra} but at variance with some others \cite{sartoris}. The relationship of this with the 
strong energy condition was also pointed out. We also pointed out a caveat in our analysis, namely that the theory becomes somewhat unphysical below a certain small radial distance, and
our analysis is strictly valid above this. However, with an appropriate choice of parameter in the $f(R)$ theory, this cutoff distance can be made very small compared to the galactic scale. 
We now move to an analysis of BSTs in the context of Brans-Dicke theory.

\section{BSTs in Brans-Dicke Theory}

We proceed to study BSTs in the formalism of the scalar-tensor Brans-Dicke theory. First let us recapitulate some basic formalism and we 
refer the reader to \cite{BDbook} for more details. 
The Brans-Dicke Lagrangian in 4-d curved space-time can be written as
\begin{equation}
L_{BD} = \sqrt{-g}\left(\varphi R - \omega\frac{1}{\varphi}g^{\mu\nu}\partial_{\mu}\varphi\partial_{\nu}\varphi+L_{\rm matter}\right),
\label{BD4d}
\end{equation}
where the real scalar-field $\varphi$ is decoupled from $L_{matter}$ and $\omega$ is the only dimensionless free parameter in this theory. 
Here we set $\hbar =c=1$ , so the mass dimension of 
$\varphi$ is $2$ and that of the gravitational constant $G_{N}$ is $-2$. In this theory, the Newtonian gravitational constant varies (as in $f(R)$ theories), and it depends on $\varphi$, 
which is a function of space-time. This latter relation can be written as 
\begin{equation}
G_{N} = \frac{1}{16\pi\varphi}.
\label{GN}
\end{equation}
Before we proceed, a few words about the relationship between the Brans-Dicke theory and the $f(R)$ models considered in the previous subsection are in order.
Recall that in the $f(R)$ theory paradigm, we wrote the action as (Eq.(\ref{fract})) as 
\begin{eqnarray}
S=\int d^4x \sqrt{-g}\,\tilde{f}(R) + S_{\rm Mat}\,,
\nonumber
\end{eqnarray}
Here we have used $\tilde{f}(R)\equiv f(R)/2\kappa$ so that the gravitational constant is absorbed by the Lagrangian as it happens in Brans-Dicke theories.
One can introduce a new field $\chi$ and write the above action as
\begin{eqnarray}
S=\int d^4x \sqrt{-g}\,\left[f(\chi) + \tilde{f}'(\chi)(R-\chi)
\right] + S_{\rm Mat}\,,
\label{newac}
\end{eqnarray}
where the prime designates a derivative with respect to the field $\chi$.  Variation with respect to $\chi$ leads to
\begin{eqnarray}
\tilde{f}''(\chi)(R-\chi)=0\,.
\label{fdprime}
\end{eqnarray}
This leads to the conclusion that $\chi=R$ if $f''(\chi)\ne 0$ as this reproduces the basic $f(R)$ action with which we started. Redefining
the field $\chi$ as $\varphi=\tilde{f}'(\chi)\equiv d\tilde{f}/d\chi$ and setting
\begin{eqnarray}
V(\phi)= \chi(\varphi) \varphi - \tilde{f}(\chi(\varphi))\,,
\label{potbd}
\end{eqnarray}
one can write the action in Eq.~(\ref{newac}) as
\begin{eqnarray}
S=\int d^4x \sqrt{-g}\,\left[\varphi R - V(\varphi)\right] + S_{\rm Mat}\,,
\label{bdact}
\end{eqnarray}
which is the Jordan frame action of a Brans-Dicke like theory of Eq.(\ref{BD4d}) with the Brans-Dicke parameter $\omega=0$, but where the 
scalar field has a potential. For our case we have $\tilde{f}(R)=\frac{1}{2\kappa}(R+\lambda
R^2)$ and consequently the potential of the scalar field turns out to be
\begin{eqnarray}
V(\varphi)= \frac{1}{8\lambda \kappa}(2\kappa \varphi -1)^2\,.
\label{vphiexp}
\end{eqnarray}
The above action and the potential specify the relationship of $f(R)$ gravity and Brans-Dicke theory in the Jordan frame. 
This correspondence is however very limited, as it only holds for $\omega=0$. To understand the full nature of the Brans-Dicke
theory, where one has BST as the solution requires to be seen explicitly for generic values of the Brans-Dicke parameter. This is
the task that we undertake now. 

Here, we closely follow the notations and conventions of \cite{BDbook}, and write this Lagrangian in a slightly different form. This is necessitated by the fact that the second 
term in the Lagrangian has a singularity when $\varphi$ becomes zero. To get rid of this singularity we set \footnote{The field $\phi$ appearing in this section is distinct from
(and should not be confused with) that appearing in section 3.}
\begin{equation}
\varphi = \frac{1}{2}\xi\phi^2,
\end{equation}
so that the new form of Lagrangian in terms of $\phi$ is 
\begin{equation}
L_{BD} = \sqrt{-g}\left(\frac{1}{2}\xi\phi^2R - \frac{1}{2}\epsilon g^{\mu\nu}\partial_{\mu}\phi\partial_{\nu}\phi +L_{\rm matter}\right),
\label{lbd}
\end{equation}
where $\epsilon = 4\omega\xi$. $\epsilon$ can take values $0,~\pm 1$, \cite{BDbook} but here we will only deal with $\epsilon = 1$.
Now if we vary $L_{BD}$ with respect to $g_{\mu\nu}$ we get 
\begin{equation}
2\varphi G_{\mu\nu} =( T_{\mu\nu})_{\rm matter}+(\partial_{\mu}\phi\partial_{\nu}\phi - \frac{1}{2}g_{\mu\nu}g^{\alpha\beta}\partial_{\alpha}\phi\partial_{\beta}\phi)
+2(\nabla_{\mu}\nabla_{\nu}-g_{\mu\nu}g^{\alpha\beta}\nabla_{\alpha}\nabla_{\beta})\varphi.
\label{BDmatter1}
\end{equation}
Similarly, by varying $L_{BD}$ with respect to $\phi$, we get 
\begin{equation}
\square\varphi = \frac{1}{2(3+2\omega)}T = \frac{\xi}{6\xi+1}T,~~~T = g^{\mu\nu}(T_{\mu\nu})_{\rm matter}.
\label{BDmatter2}
\end{equation}
For our spherically symmetric static BST, we choose $\phi =\phi(r)$, and then we get using Eq.(\ref{BDmatter1}), \footnote{The quantities $\rho$, $P_i$ refer to the matter part of the Lagrangian
of Eq.(\ref{lbd}), and is obtained from the matter part of Eq.(\ref{BDmatter1}). This will be understood in what follows, and we will avoid using a subscript, as this clutters up the notation.}
\begin{eqnarray}
\rho &=& -\frac{\beta^{2}(4\xi+1)}{2}\left(\frac{d\phi}{dr}\right)^{2}-4\beta^2\xi\frac{\phi}{r}\left(\frac{d\phi}{dr}\right)-2\beta^2\xi\phi\left(\frac{d^2\phi}{dr^2}\right)+\xi\phi^2\frac{1-\beta^2}{r^2}\nonumber\\
P_{r} &=& -\frac{\beta^2}{2}\left(\frac{d\phi}{dr}\right)^2 +\frac{(4r+5r_{b})}{r(r+r_{b})}\beta^2\xi\phi\left(\frac{d\phi}{dr}\right)-\left(\frac{1}{r^2}-\frac{(r+2r_{b})}{r^2(r+r_{b})}\right)\xi\phi^2\nonumber\\
P_{\theta} &=& P_{\phi}=\frac{(4\xi+1)\beta^2}{2}\left(\frac{d\phi}{dr}\right)^2+\frac{(2r+3r_{b})}{r(r+r_{b})}\beta^2\xi\phi\left(\frac{d\phi}{dr}\right) + 2\beta^2\xi\phi\left(\frac{d^2\phi}{dr^2}\right)\nonumber\\
&+&\xi\phi^2\frac{r_{b}(r_{b}-2r)\beta^2}{4r^2(r+r_{b})^2},
\end{eqnarray}
where $r_{b}=\frac{\alpha}{D}$, and we have defined as usual, $T^0_0 = -\rho$, $T^1_1=P_r$, $T^2_2=P_{\theta}$, $T^3_3=P_{\phi}$.
Also, from Eq.(\ref{BDmatter2}), we get :
\begin{equation}
\beta^2\xi\frac{4r+5r_{b}}{2(r+r_{b})}\frac{\phi}{r}\left(\frac{d\phi}{dr}\right)+\beta^2\xi\phi\left(\frac{d^2\phi}{dr^2}\right)+\xi\beta^2\left(\frac{d\phi}{dr}\right)^2 = \frac{\xi}{6\xi+1}T\,.
\label{BDmatter3}
\end{equation}
Now from Eq.(\ref{BDmatter3}), by substituting for $T$ from Eq.(\ref{BDmatter1}), we obtain the following linear differential equation of $\phi$ 
\begin{equation}
\frac{d^2\phi}{dr^2}+\frac{4r+5r_{b}}{2(r+r_{b})r}\left(\frac{d\phi}{dr}\right)+\left(\frac{4}{2r^2\beta^2}-\frac{(4r^2+8rr_{b}+7r_{b}^2)}{2(r+r_{b})^2r^2}\right)\xi\phi=0
\label{demain}
\end{equation}
The general solution of Eq.(\ref{demain}) is difficult to obtain analytically, and we will momentarily study numerical solutions. However, it is instructive to first look at some simple limits. 
First, let us set $\xi = 0$, in which case Eq.(\ref{demain}) and its solution with arbitrary constants $C_1$ and $C_2$ is
\begin{equation}
\frac{d^2\phi}{dr^2}+\frac{4r+5r_{b}}{2(r+r_{b})r}\left(\frac{d\phi}{dr}\right)=0,~~\implies 
\phi = -\frac{2(r+r_b)^{3/2}}{3r^{3/2}r_b}C_1 + C_2.
\end{equation}
Now on physical grounds, if we demand the solution to be regular near the origin, then we need to set $C_1 = 0$, in which case $\phi$ is a constant as expected, since in the 
limit $\xi =0$, the Brans-Dicke theory goes over to GR, where $G_N$ has a fixed value. 
\begin{figure}[t!]
\begin{minipage}[b]{0.5\linewidth}
\centering
\includegraphics[width=2.7in,height=2.3in]{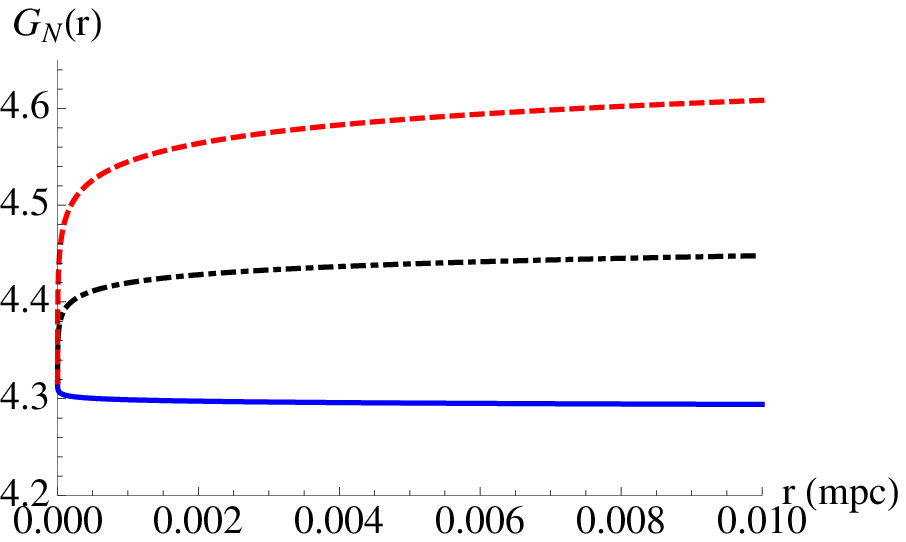}
\caption{Effective Newton's constant as a function of distance for BSTs in Brans-Dicke theory (see text for details).}
\label{gn}
\end{minipage}
\hspace{0.2cm}
\begin{minipage}[b]{0.5\linewidth}
\centering
\includegraphics[width=2.8in,height=2.3in]{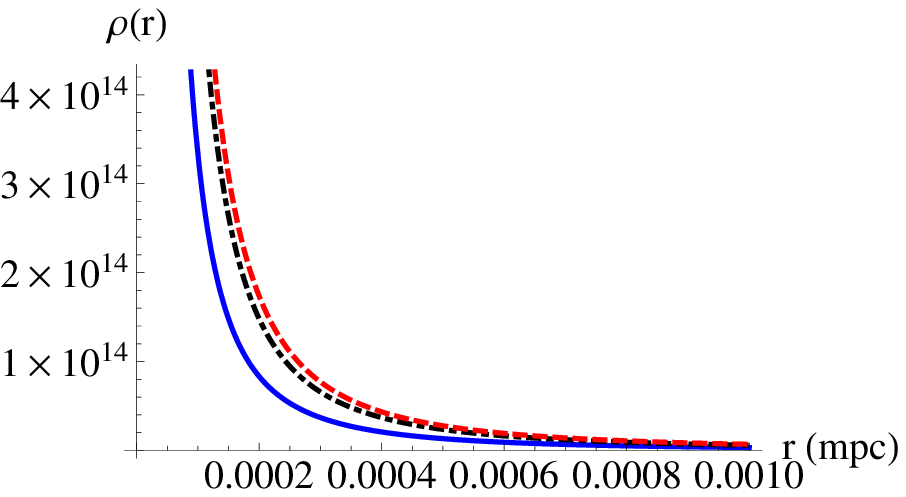}
\caption{Matter density as a function of distance for BSTs in Brans-Dicke theory (see text for details).}
\label{mat1}
\end{minipage}
\end{figure}
We then take the limit $r \ll r_b$, in which case Eq.(\ref{demain}) reduces to
\begin{equation}
\frac{d^2\phi}{dr^2} + \frac{5}{2r}\left(\frac{d\phi}{dr}\right) + \frac{1}{2r^2}\left(\frac{4-7\beta^2}{\beta^2}\right)\xi\phi=0,
\end{equation}
which leads to a power law solution 
\begin{equation}
\phi(r)^{\rm near} = C_3r^{a_+} + C_4r^{a_-},~~~a_{\pm} = -\frac{3 \beta \pm i \sqrt{32 \xi -\beta ^2 (56 \xi + 9)}}{4 \beta},
\label{solnear}
\end{equation}
where $C_3$ and $C_4$ are arbitrary constants, to be fixed from boundary conditions. Similarly, in the limit $r \gg r_b$, Eq.(\ref{demain}) becomes
\begin{equation}
\frac{d^2\phi}{dr^2} + \frac{2}{r}\left(\frac{d\phi}{dr}\right) + \frac{2}{2r^2}\left(\frac{1-\beta^2}{\beta^2}\right)\xi\phi=0,
\end{equation}
and yields the solution, with arbitrary constants $C_5$  and $C_6$
\begin{equation}
\phi(r)^{\rm far} = C_5r^{b_+} + C_6r^{b_-},~~~b_{\pm} = -\frac{\beta \pm i \sqrt{8 \xi -\beta ^2 (8 \xi +1)}}{2 \beta}.
\label{solfar}
\end{equation}
If we demand that the near and far solutions be real (we had started with a real scalar field), then Eqs.(\ref{solnear}) and (\ref{solfar}) give the constraints 
\begin{eqnarray}
&~& {\rm Near~region~:} ~~~0<\beta<\frac{2}{\sqrt{7}},~\xi < \frac{9\beta^2}{8(4-7\beta^2)},~~{\rm or}~~\beta>\frac{2}{\sqrt{7}},~\xi > \frac{9\beta^2}{8(4-7\beta^2)} \nonumber\\
&~& {\rm Far~region~:} ~~~0<\beta<1,~\xi < \frac{\beta^2}{8(1-\beta^2)}.
\label{constraint}
\end{eqnarray}
The second relation of Eq.(\ref{constraint}) is merely the statement that $\xi$ should be taken as positive, and should not be thought of as a lower bound on $\xi$. 
We also remind the reader that in the original BST of Eq.(\ref{type2a}), we must necessarily have $0<\beta<1$. Now remembering that the Brans-Dicke parameter is defined by $\omega = 1/(4\xi)$, 
we find that in the near region, for $\beta < 2/\sqrt{7}$, $\omega$ is constrained to be greater than $2(4-7\beta^2)/(9\beta^2)$. In the far region, $\omega > 2(1-\beta^2)/\beta^2$.

We now comment on the general solution to Eq.(\ref{demain}). We will choose $\xi = 10^{-3}$ and $\beta =  0.8,~0.6,~0.5$ for illustration. From Eq.(\ref{GN}), we have that $G_{N} = 1/(8\pi\xi\phi^2)$, and
the boundary conditions on $\phi$ follows from this, and the nature of the solution is entirely dependent on the boundary conditions. We choose $\phi = 9.6 \times 10^3$ and $\phi' = 0.1$ at
$r = 10^{-7}$, and numerically solve Eq.(\ref{demain}). In Fig.(\ref{gn}), we plot the effective Newton's constant as a function of $r$. The solid blue, dotted black and dashed red lines 
correspond to $\beta =  0.8,~0.6,~0.5$ respectively, and the $y$ axis is scaled by a factor of $10^{-9}$. It is seen that depending on the value of $\beta$, $G_N(r)$ becomes effectively constant
close to the origin. It is also important to check that the matter energy density is positive in our numerical scheme. This is shown in Fig.(\ref{mat1}), where the same color coding as in Fig.(\ref{gn}) has been used. 

We now summarize the results of this section. Here, we have considered Bertrand space-times in the Brans-Dicke theory of gravity. We obtained the differential equation for the
Brans-Dicke scalar $\phi$ and obtained analytical solutions in some simple limits. These were shown to give rise to interesting constraints on the BST parameter $\beta$ and the Brans-Dicke
parameter $\xi$. Further, we computed $\phi$ numerically and checked that the matter energy density is positive definite, as is required for a physical theory. 

\section{Conclusions and Discussions}

The results of this paper strengthen the arguments made in our earlier works \cite{dbs1},\cite{dbs2}. Broadly, in this paper, we have shown that 
Bertrand space-times provide a viable model for galactic dark matter, even in extended theories of gravity. 

This article starts with the  delicate nature of the definition of circular velocity of stars in a spherically symmetric, static space-time in general relativity. In section 2 we have shown 
that there can be, in general, two different ways in which the circular velocities of the stars are defined. In one way the velocity is measured by a local observer situated near the star and
in the other, no such requirement is necessary. The discussion on these definitions shows that as far as velocity rotation curves are concerned, the latter definition makes more sense 
because in the light of this definition of stellar velocity, one  can compare the velocities of stars at different radial distances from the core of the galaxy. In this context,
we also provided a spectroscopic interpretation of the formula for the galactic rotation curve for BST observers. 

In section 3, we showed that a real scalar field or a radiation field cannot seed a BST. 
Next, we extended BSTs, previously studied in a general relativistic framework, to the realm of modified gravity theories.  
In this regard, note that BSTs are interesting when one looks for stable and closed circular
geodesics. In GR, it can be shown that BST's cannot exist without matter. Consequently, the next question rises which compels one to
search for modified gravity situations where it may happen that BST's may exist without explicit hydrodynamic matter. In our analysis we
showed that that in the $f(R)$ paradigm and in Brans-Dicke theories, we do not get BST solutions without matter. This result does confirm some realistic
situations, where, from the Bullet cluster results \cite{Clowe}, it is now accepted that some form of dark matter does exist.

Next, we studied BSTs in $f(R)$ and Brans-Dicke theories. 
It should be remembered that unlike the works of \cite{Capo2}, \cite{Capo3}, in this paper we did not solve for the metric in $f(R)$ or Brans-Dicke theories. 
The BST was assumed as a solution for these. We saw that if one uses the BST as a solution, then one cannot neglect the contribution of matter 
(in our case dark matter) for the solutions in $f(R)$ theory or Brans-Dicke theory (with arbitrary $\omega$).
In section 4, in the context of $f(R)$ theories,  we pointed out various aspects of the Newtonian potential, and further analyzed the equation of state parameter 
in $f(R)$ gravity. In section 5, we established the nature of BSTs in Brans-Dicke theories, and showed that these might indicate 
interesting constraints on the Brans-Dicke as well as the BST parameters. The physics of these constrains should be interesting to study further.

\begin{center}
{\bf Acknowledgements}
\end{center}
It is a pleasure to thank Sayan Kar for valuable comments.

\end{document}